\pdfoutput=1
\documentclass[a4paper,11pt]{article}
\usepackage{jheppub} %
\usepackage{lineno}
\usepackage{color}
\usepackage{graphicx}
\usepackage{float}
\usepackage[usenames,dvipsnames]{xcolor}
\usepackage{multirow}
\usepackage{braket}
\usepackage{booktabs}

\providecommand{\Ssteer}{\mathcal{S}}            %
\providecommand{\Mtt}{M_{t\bar t}}
\providecommand{\ttbar}{t\bar t}

\usepackage[normalem]{ulem}

\title{\boldmath Steerability in top quark decays at the LHC
}

\author[a]{Tao Han,}
\author[b,c,d]{Navin McGinnis,}
\author[b]{Sepehr Salamat,}
\author[b]{Shufang Su}
\author[b]{and Xingzhou Yu}

\affiliation[a]{PITT PACC, Department of Physics and Astronomy,\\ University of Pittsburgh, 3941 O’Hara St., Pittsburgh, PA 15260, USA}
\affiliation[b]{Department of Physics, University of Arizona, Tucson, Arizona 85721, USA}
\affiliation[c]{Enrico Fermi Institute, Physics Department, University of Chicago, Chicago, IL 60637, USA}
\affiliation[d]{High Energy Physics Division, Argonne National Laboratory, Lemont, IL 60439, USA}

\emailAdd{than@pitt.edu}
\emailAdd{nmcginnis@uchicago.edu}
\emailAdd{sepehrsalamat@arizona.edu}
\emailAdd{shufang@arizona.edu}
\emailAdd{xingzhou@arizona.edu}

\abstract{
Quantum information observables in top quark pair production have opened a new experimental window on non-classical correlations of particle interactions at colliders. Einstein-Podolsky-Rosen (EPR) steering occupies a distinguished place in the quantum hierarchy, lying between entanglement and Bell nonlocality. We present a detailed collider analysis of steerability for top quark pairs at the Large Hadron Collider (LHC). In particular, we utilize both the decay and kinematic methods to estimate the expected sensitivity of the LHC and High Luminosity LHC (HL-LHC) to detect steerable states. We find that with the current LHC dataset of $139~\mathrm{fb}^{-1}$ the steerability of the $t\bar t$ state in the boosted region can be established at $7.7\sigma$ ($1.8\sigma$) in the semileptonic (dileptonic) channel with the decay method, corresponding to a measurement of the steerability integral at a precision of $1.5\%$ ($5.2\%$), while the kinematic method reaches a precision of $0.05\%$ in both channels. At the HL-LHC the decay-method precision improves to $0.3\%$ ($1.1\%$). Near threshold, where the inclusive selections used for entanglement and discord yield a state at the steerability boundary, an optimized selection $\Mtt\lesssim360$--$400$~GeV isolates a steerable singlet-dominated state, where the sensitivity is projected to reach $7\sigma$ ($3.4\sigma$). We also provide a detailed discussion on the interpretation of these results with respect to fictitious quantum states, and argue for the robust nature of the steerability measurement at colliders.}

\begin{document}
\preprint{PITT-PACC-2623}
\maketitle
\flushbottom

\section{Introduction}
\label{sec:intro}

Experiments at the Large Hadron Collider (LHC) have established that the spin quantum numbers of pair produced particles form experimentally accessible bipartite quantum states, and that non-classical correlations in these states can be reconstructed and measured. The observation of quantum entanglement in top quark pairs by the ATLAS~\cite{ATLAS:2023fsd} and CMS~\cite{CMS:2024pts} collaborations has turned the top quark into an ideal laboratory for quantum information, and incited a broad body of proposals for entanglement, Bell variables and quantum tomography at high-energy colliders: for top quarks and other fermion pairs~\cite{Afik:2020onf,Fabbrichesi:2021npl,Severi:2021cnj,Aoude:2022imd,Fabbrichesi:2022ovb,Afik:2022dgh,Ashby-Pickering:2022umy,Severi:2022qjy,Altakach:2022ywa,Dong:2023xiw,Morales:2023gow,Aoude:2023hxv,Ma:2023yvd,Sakurai:2023nsc,Bernal:2023jba,Han:2023fci,Altomonte:2023mug,Ehataht:2023zzt,Maltoni:2024tul,Aguilar-Saavedra:2024hwd,Blasone:2024dud,Subba:2024mnl,Bernal:2024xhm,Wu:2024asu,Demina:2024dst,Gabrielli:2024kbz,Ruzi:2024cbt,CMS:2024zkc,Du:2024sly,Ravina:2024ard,Cheng:2024rxi,Sullivan:2024wzl,Wu:2024ovc,Ruzi:2024iqu,Altomonte:2024upf,Han:2024ugl,Fabbrichesi:2025ywl,Cheng:2025cuv,Lysak:2025uhk,Han:2025ewp,Fabbrichesi:2025aqp,Aoude:2025ovu,Fabbrichesi:2025psr,Aguilar-Saavedra:2025byk,Aoude:2025jzc,Afik:2025grr,Cheng:2025zcf,Bechtle:2025ugc,Abel:2025skj,Wu:2025dds,Aguilar-Saavedra:2025cej,Pei:2025ito,Yazgan:2025pah,Cao:2025xnp,CMS:2025brx,Jolly:2026gpe,Pardos:2026uhw,Guo:2026yhz,Gabrielli:2026tnl,Zhang:2026nwm,Yang:2026uwu,Aguilar-Saavedra:2026wuq,Oussarhan:2026yli,Fang:2026ddi,Aoude:2026eeg,Arai:2026jtc,ATLAS:2026nrx,Subba:2026nzs,Wang:2026nls,Zhang:2026wvn,Zhou:2026poo,Goncalves:2026nnx,Liu:2026gxj,Batell:2026bcd,Lamba:2026sni,Lamba:2026jfm,deLima:2026jbb,Erdogan:2026cbe,A:2026pug,Belmonte:2026yno,Antozzi:2026vdi,Martinez-Moreno:2025jgx,Choi:2026omc,Altakach:2026fpl,Afik:2026api}, for gauge and Higgs bosons~\cite{Goncalves:2025mvl,Ruzi:2025jql,Ding:2025mzj,Cheng:2026ktp,Lamba:2026qnk,Liu:2025iwh,Banacki:2026msu,Goncalves:2026njf,Goncalves:2025xer,DelGratta:2025xjp,Pelliccioli:2026ltl,Haisch:2025jqr}, where entanglement has since been measured as well~\cite{ATLAS:2026hye}, for hadronic, electron--ion and QED systems~\cite{Hatta:2024lbw,Kanachova:2026ywy,Feng:2025ryr,Sharma:2026ymt,Joshi:2026zlb,Xie:2026vvs,Yang:2026rye,Cao:2026zqo,Fucilla:2026zhy,Fucilla:2026mkg,Fucilla:2025kit,Cheng:2025zaw,Qi:2025onf,Hentschinski:2026otq,Kutak:2025awi,Hatta:2025obw,Zhang:2025ean,Qian:2026aoh,Hatta:2026dqs}, and on the foundations and tools of such measurements~\cite{Low:2025aqq,Aguilar-Saavedra:2026rsx,Dudal:2024bmf,Bernal:2024dtg,Caban:2026oih,Clarke:2026vwr,Liu:2026dzv,Gu:2025ijz,Blasone:2025lus,McGinnis:2025xgt,Durupt:2025wuk,ATLAS:2026bre}; see Refs.~\cite{Barr:2024djo,Afik:2026iyl} for reviews.

The space of quantum correlations is much richer than entanglement, and measures beyond entanglement and Bell nonlocality, such as quantum discord, the nonlocal advantage of quantum coherence~\cite{Afik:2026api} and nonstabilizerness or ``magic''~\cite{Li:2026zum,Gargalionis:2025iqs,Gargalionis:2026onv,Viaux:2026skf,Alvarado:2026owd}, have begun to be explored at colliders. In particular, there are distinct notions of correlations which form a hierarchy based on the amount of information that can be gained through projective measurements: Bell-nonlocal states $\subset$ steerable states $\subset$ entangled states $\subset$ states with non-zero quantum discord. Quantum discord for top quark pairs was proposed in Ref.~\cite{Afik:2022dgh} and studied in detail for the LHC in Ref.~\cite{Han:2024ugl}, and, together with steering, in $e^+e^-$ annihilation with polarized beams in Ref.~\cite{Zhang:2026wvn}. In this work, we turn to Einstein-Podolsky-Rosen (EPR) steering following a similar strategy of collider analysis.

Quantum steering formalizes the original EPR scenario as a
one-sided task: Alice, by her choice of local measurement,
remotely conditions the ensemble of states held
by Bob in a way that cannot be reproduced by any
local hidden state (LHS) model. Two features make
steering qualitatively different from the correlations
studied so far at colliders. First, it is intrinsically
asymmetric: whether Alice can steer Bob does not
necessarily coincide with whether Bob can steer Alice. Second,
it is strictly intermediate between entanglement and
Bell nonlocality, {\it i.e.} there exist entangled states that are
not steerable, and steerable states that do not violate
any Bell inequality~\cite{Wiseman:2007hyt}. Measuring steerability at
the LHC therefore probes a sharper notion of quantum
correlation than entanglement, while remaining accessible
in regions of phase space where Bell violation is
out of reach; for the status of Bell tests at colliders as tests of local
realism see Refs.~\cite{Low:2025aqq,Aguilar-Saavedra:2026rsx}.

The possibility of accessing steerable quantum states in top quark pairs was first presented in Ref.~\cite{Afik:2022dgh}. Recently, a geometric framework based on quantum steering ellipsoids~\cite{Jevtic:2014icx} was formulated for collider states in Ref.~\cite{Alvarez:2026uve}, characterizing the geometry of the reconstructed density matrix, and the hierarchy of quantum correlations in top quark pairs has been examined on the basis of the existing measurements in Ref.~\cite{Afik:2026pxv}. In this work, we study steerability as in Ref.~\cite{Afik:2022dgh} across the phase space of top quark pairs providing optimized kinematic cuts to enhance the sensitivity at the LHC. We also provide closed-form steering criteria for the leading order cross section for top quark pairs. We perform a detailed collider analysis of the semileptonic and dileptonic decay channels, including the full simulation of hadronization, and detector simulation. For the reconstructed kinematic distributions we also perform a detailed unfolding procedure to estimate the systematic error expected at the LHC. Further, we present the results using two methods for reconstructing quantum states at colliders: the decay method and the kinematic method~\cite{Han:2024ugl,Cheng:2024rxi}.

Using the decay method, we find that the steerability of the boosted $t\bar{t}$ state can be established with current data at $7.7\sigma$ in the semileptonic channel and at $1.8\sigma$ in the dileptonic channel, corresponding to a precision of $1.5\%$ and $5.2\%$ in the steerability integral; at the end of the high luminosity LHC (HL-LHC) these improve to $0.3\%$ and $1.1\%$, respectively. The kinematic method improves precision by one to two orders of magnitude, reaching $0.05\%$ in both channels with current data. Finally, we develop the necessary formalism to provide the first interpretation of steering for the fictitious states reconstructed at colliders. We show that the steerability integral is convex in the spin correlation matrix, so that a steerable fictitious state certifies steerable $t\bar t$ states in the selected region of phase space. We thus establish robust criterion for a genuine measurement of steerability at particle colliders.

We note that our treatment is within the framework of Quantum Field Theory (QFT) and in the Standard Model (SM) of particle physics. Accordingly, the measurements in high energy physics experiments by constructing particle decays explore the intrinsic quantum nature of  entanglement and test the self-consistency of quantum theory in the unprecedented kinematic regime. It should not be interpreted as to exclude a generic Local Hidden Variable Theory (LHVT) \cite{Abel:1992kz,Low:2025aqq,Abel:2025skj} in the sense of the EPR paradox \cite{Einstein:1935rr}, although a specific LHVT could be falsified and thus excluded \cite{Go_2007}.

This paper is organized as follows. We review the description of $t\bar{t}$ spin-correlations at the LHC and their information-theoretic interpretation in terms of bipartite states of qubits Sec.~\ref{sec:quantum_states}. In Sec.~\ref{sec:steerability}, we introduce the concept of quantum steerability in the context of collider observables, and provide a discussion of its evaluation in terms of fictitious states. In Sec.~\ref{sec:collider_analysis}, we perform a detailed collider analysis of spin-correlations in the $t\bar{t}$ system and provide current and projected measurement prospects for steerability. We conclude in Sec.~\ref{sec:conclusions}.

\section{Quantum states of $t\bar{t}$ at the LHC}
\label{sec:quantum_states}

\subsection{Spin density matrix}
\label{sec:density_matrix}

We consider top quark pairs ($t\bar{t}$) produced at the LHC, whose spin
degrees of freedom form a two-qubit system. The quantum state is described by
the density matrix in the Fano--Bloch form~\cite{Fano:1957zz,Fano:1983zz}
\begin{equation}
\rho_{t\bar{t}}=\frac{1}{4}\left(\mathbb{I}\otimes\mathbb{I}
+ \sum_i B^{+}_{i}\sigma^{i}\otimes\mathbb{I}
+ \sum_{j}B^{-}_{j}\mathbb{I}\otimes\sigma^{j}
+ \sum_{ij}C_{ij}\sigma^{i}\otimes\sigma^{j}\right),
\label{eq:fano}
\end{equation}
where $\mathbb{I}$ is the $2\times2$ identity matrix and $\sigma^{i}$,
$i=1,2,3$, are the Pauli matrices. The vectors $B^{\pm}_{i}$ are the
polarizations of $t$ and $\bar{t}$ and $C_{ij}$ is the spin correlation
matrix. The spins are measured along a set of quantization axes. We use the
\textit{helicity} basis $\{\hat k,\hat r,\hat n\}$, defined in the $t\bar t$
center-of-mass (COM) frame with $\hat k$ along the top-quark momentum,
$\hat r=(\hat z-\cos\theta\,\hat k)/\sin\theta$ the component of the beam
direction $\hat z$ orthogonal to $\hat k$, and $\hat n=\hat r\times\hat k$
normal to the production plane, where $\cos\theta=\hat k\cdot\hat z$ defines
the scattering angle $\theta$; the same axes are used for $t$ and
$\bar t$~\cite{Han:2023fci}. We also refer to the \textit{beam} basis
$\{\hat x,\hat y,\hat z\}$, fixed by the beam direction and two transverse
axes. The two bases coincide, up to a relabeling of the axes, in the
transverse configuration $\theta=\pi/2$.

The symmetries of QCD constrain the coefficients in
Eq.~\eqref{eq:fano}~\cite{Bernreuther:2015yna}. In particular, parity invariance forbids
polarizations in the production plane, $B^{\pm}_{k}=B^{\pm}_{r}=0$, and
the correlation coefficients
$C_{kn}=C_{rn}=0$. CP invariance imposes $B^{+}_{i}=B^{-}_{i}$ and
$C_{ij}=C_{ji}$. The normal polarization $B_n$, which parity allows but which
is odd under naive time reversal, requires absorptive parts and first appears
at one loop in QCD, while the in-plane polarizations arise from
parity-violating weak corrections; both are at the level of a few $10^{-3}$ in
the SM~\cite{Bernreuther:2015yna}, and we neglect them together with the
electroweak corrections to the spin correlations~\cite{Frederix:2021zsh}. At
leading order the state is therefore characterized by $B^{\pm}=0$ and by a
real symmetric correlation matrix with four non-vanishing entries, $C_{kk}$,
$C_{rr}$, $C_{nn}$ and $C_{kr}=C_{rk}$.

A two-qubit state with vanishing polarizations is Bell diagonal in the basis
that diagonalizes $C$. If $c_1,c_2,c_3$ are the eigenvalues of $C$, the
eigenvalues of $\rho_{t\bar{t}}$ are
\begin{equation}
\lambda_{B} = \frac14\Big(1+\sum_i c_i\,c^{(B)}_i\Big),\qquad
c^{(B)}\in\big\{(-1,-1,-1),\,(1,1,-1),\,(1,-1,1),\,(-1,1,1)\big\},
\label{eq:belleig}
\end{equation}
giving the weights of the four Bell states $B$, with correlation vectors $c^{(B)}$.  Positivity of the density matrix confines $(c_1,c_2,c_3)$ to a
tetrahedron with these four vertices, and the separable states form the
octahedron $|c_1|+|c_2|+|c_3|\le1$~\cite{Horodecki:1996nc}. Every quantum-information property of the
leading-order $t\bar t$ state is then a
function of the position of the state in this tetrahedron.

\subsection{The leading-order state over phase space}
\label{sec:lo_state}

At leading order, $t\bar t$ pairs are produced by $q\bar q$ annihilation and
$gg$ fusion. The spin state at a point in phase space, depending on the
scattering angle $\theta$ and the invariant mass $M_{t\bar t}$ is given by the luminosity-weighted
sum
\begin{equation}
\label{eq:rhoThetaMtt}
\rho(\theta,M_{t\bar{t}})=\frac{L_{q\bar{q}}|\mathcal{M}_{q\bar{q}\to t\bar{t}}|^{2}\rho_{q\bar{q}}(\theta,M_{t\bar{t}}) + L_{gg}|\mathcal{M}_{gg\to t\bar{t}}|^{2}\rho_{gg}(\theta,M_{t\bar{t}})}{L_{q\bar{q}}|\mathcal{M}_{q\bar{q}\to t\bar{t}}|^{2} + L_{gg}|\mathcal{M}_{gg\to t\bar{t}}|^{2}},
\end{equation}
where $L_{I}$ are the partonic luminosities and
$|\mathcal{M}_{I\to t\bar{t}}|^{2}$ the spin-averaged matrix elements for
$I=q\bar{q},gg$. Each $\rho_I$ is itself a mixed state
because the helicities of the initial partons are averaged and different
helicity configurations populate different spin states of the
pair~\cite{Mahlon:2010gw}. In $gg$ fusion, like-helicity gluons
($J_z=0$ along the beam) produce the pair in a spin singlet, which dominates
near threshold, while unlike-helicity gluons ($|J_z|=2$) produce a spin
triplet that dominates at high $p_T$. In $q\bar q$ annihilation the vector
coupling produces a triplet with $|J_z|=1$ along the beam. 

\begin{figure}[t]
\centering
\includegraphics[width=0.75\textwidth]{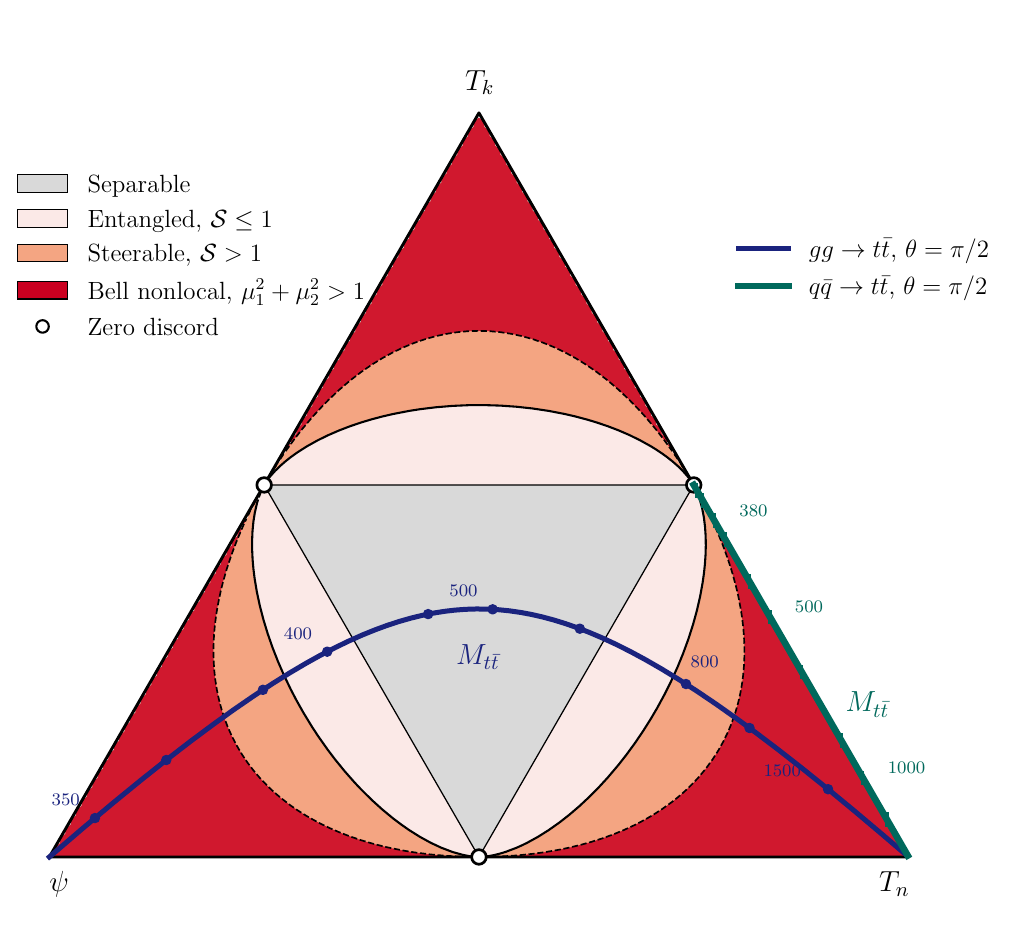}
\caption{The face of the Bell tetrahedron spanned by the singlet $\psi$ and
the triplets $T_n$ and $T_k$. Shaded are the separable states, the
entangled but unsteerable states (light red), the steerable states ($\Ssteer>1$,
Sec.~\ref{sec:steer-criterion}) and the states violating the CHSH inequality
($\mu_1^2+\mu_2^2>1$). The open circles at the edge midpoints are the only
states on the face with vanishing quantum discord. The curves are the
leading-order $gg\to t\bar t$ and $q\bar q\to t\bar t$ states at
$\theta=\pi/2$, Eqs.~\eqref{eq:rhoqq_bell}--\eqref{eq:rhogg_bell}, with
$M_{t\bar t}$ in GeV indicated along them; the physical $pp$ state is a
mixture of the two and lies between them.}
\label{fig:face}
\end{figure}

In the transverse configuration $\theta=\pi/2$, where $C_{kr}=0$ and $C_{ij}$ is
diagonal in the helicity basis, this structure can be made fully explicit.
Let $\psi$ denote the spin singlet and $T_n$, $T_k$, $T_r$ the spin-triplet
states with vanishing spin projection along $\hat n$, $\hat k$ and $\hat r$,
with correlation vectors $(c_k,c_r,c_n)=(-1,-1,-1)$, $(1,1,-1)$, $(-1,1,1)$
and $(1,-1,1)$, respectively. From the leading-order spin
correlations~\cite{Afik:2020onf,Aoude:2022imd} one finds
\begin{align}
\rho_{q\bar{q}}\big(\beta,\tfrac{\pi}{2}\big) &=
\frac{T_{n} + (1-\beta^{2})\,T_{k}}{2-\beta^{2}},
\label{eq:rhoqq_bell}\\
\rho_{gg}\big(\beta,\tfrac{\pi}{2}\big) &=
\frac{(1-\beta^{2})\,\psi + \beta^{2}\,T_{n} + 2\beta^{2}(1-\beta^{2})\,T_{k}}
{1+2\beta^{2}-2\beta^{4}}.
\label{eq:rhogg_bell}
\end{align}
where $\beta = \sqrt{1-4m_{t}^{2}/M^{2}_{t\bar{t}}}$ is the top-quark velocity.
Equations~\eqref{eq:rhoqq_bell}--\eqref{eq:rhogg_bell} are equivalent to the
decomposition into Bell states and separable mixtures given in
Refs.~\cite{Barr:2024djo,Fabbrichesi:2022ovb,Han:2024ugl}, but expose a
simple structure where both channels are mixtures of the same three Bell states
with polynomial weights, and the fourth Bell state $T_r$, the triplet with
vanishing spin projection along the beam, is never produced. The
$\theta=\pi/2$ states therefore lie on a single face of the Bell tetrahedron,
shown in Fig.~\ref{fig:face}. The $q\bar q$ state is a rank-two mixture of
$T_n$ and $T_k$ and moves along an edge of the face, from the separable
equal-weight mixture at threshold to the pure triplet $T_n$ for $\beta\to1$. The $gg$ state moves
through the interior of the face from the singlet at threshold to the same
triplet at high $p_T$, passing at $\beta^2=1/2$ ($M_{t\bar t}\simeq490$~GeV)
through the maximally mixed point of the face, where the three weights are
equal and the state is separable. The corresponding spin correlation
matrices are
\begin{align}
C_{q\bar q}\big(\beta,\tfrac{\pi}{2}\big) &= \mathrm{diag}\big(a,\,1,\,-a\big),
\qquad a=\frac{\beta^2}{2-\beta^2},
\label{eq:Cqq}\\
C_{gg}\big(\beta,\tfrac{\pi}{2}\big) &=
\frac{\mathrm{diag}\big(-1+2\beta^4,\;-1+4\beta^2-2\beta^4,\;-(1-2\beta^2+2\beta^4)\big)}
{1+2\beta^2-2\beta^4}.
\label{eq:Cgg}
\end{align}
Fig.~\ref{fig:face} also shows, on the same face, the regions of separable,
steerable and Bell-nonlocal states, anticipating the steering criterion of
Sec.~\ref{sec:steerability}; the nested structure of the hierarchy and its
interplay with the kinematics of $t\bar t$ production can be read off
directly. Away from $\theta=\pi/2$ the correlation $C_{kr}$ is non-zero and
the state leaves this face, but the qualitative picture is unchanged.

\begin{figure}
\centering
  \includegraphics[scale=0.45]{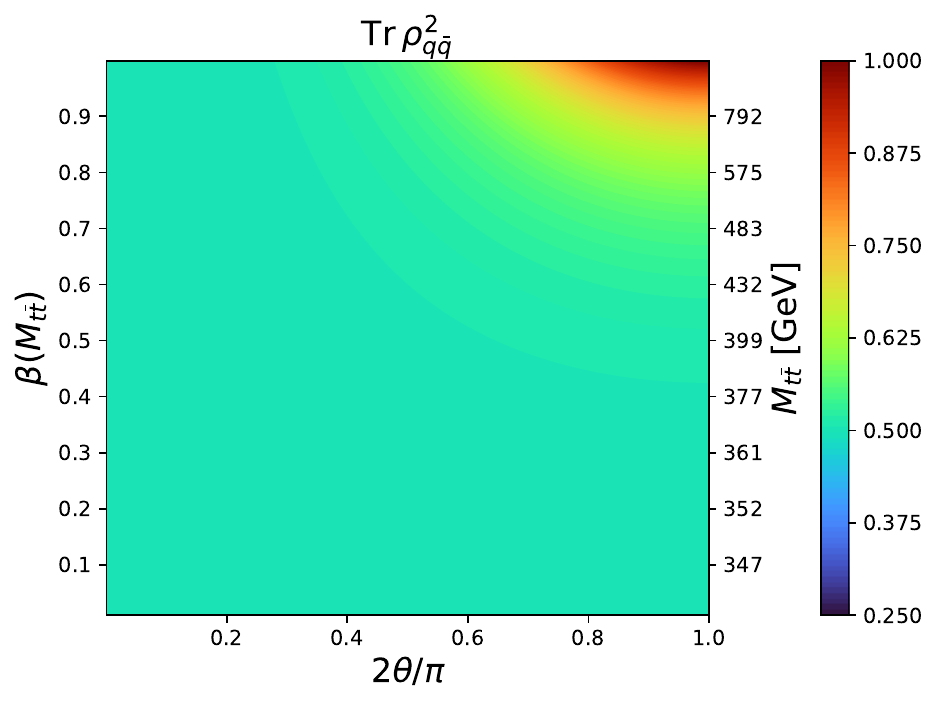}
  \includegraphics[scale=0.45]{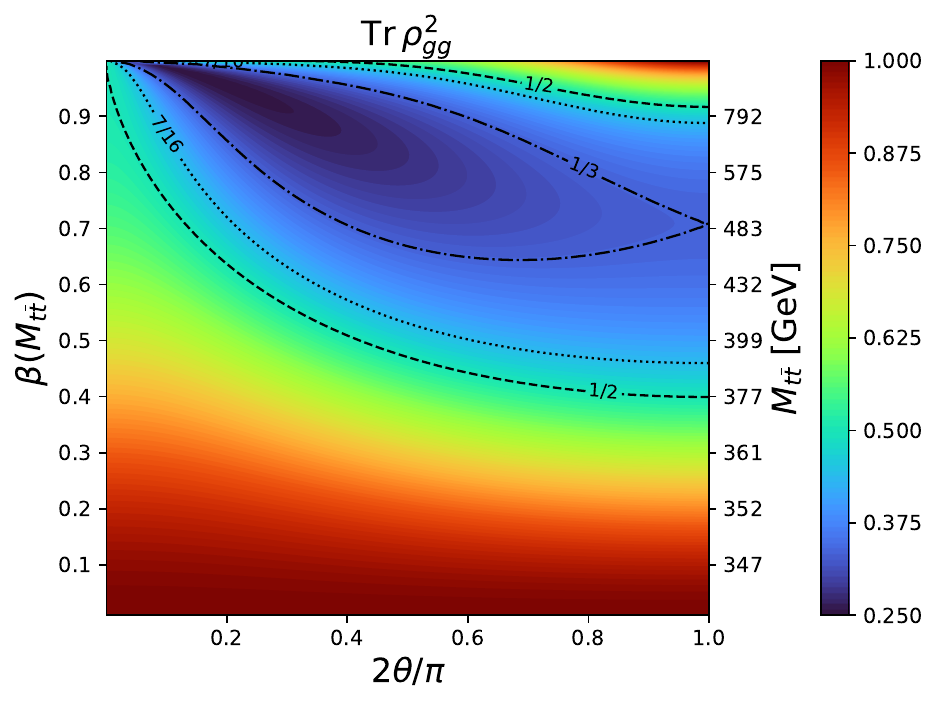}\\
  \includegraphics[scale=0.45]{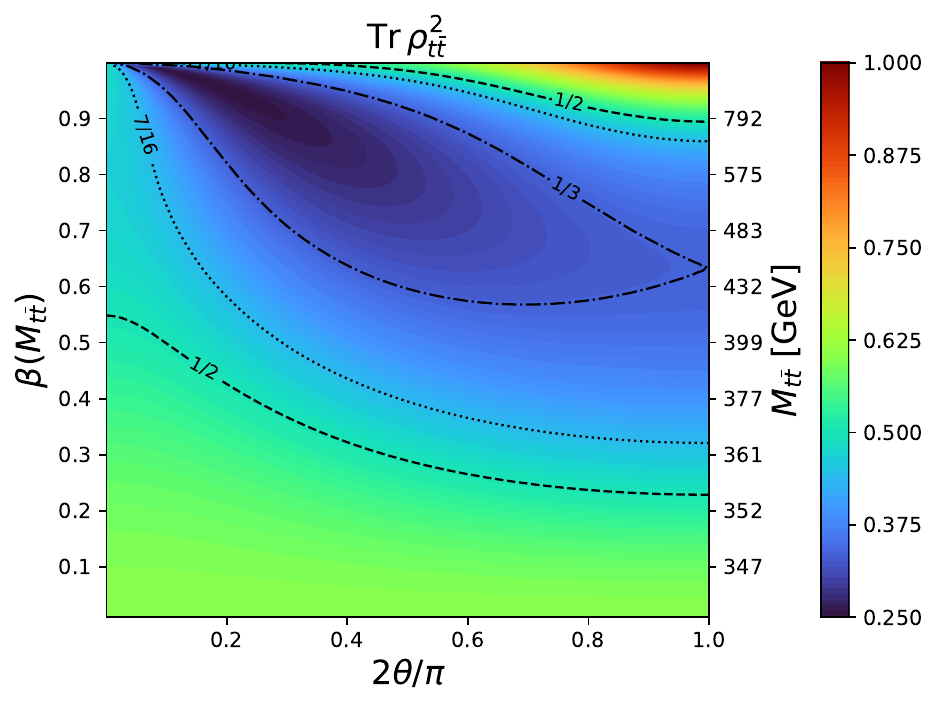}  
  \includegraphics[scale=0.45]{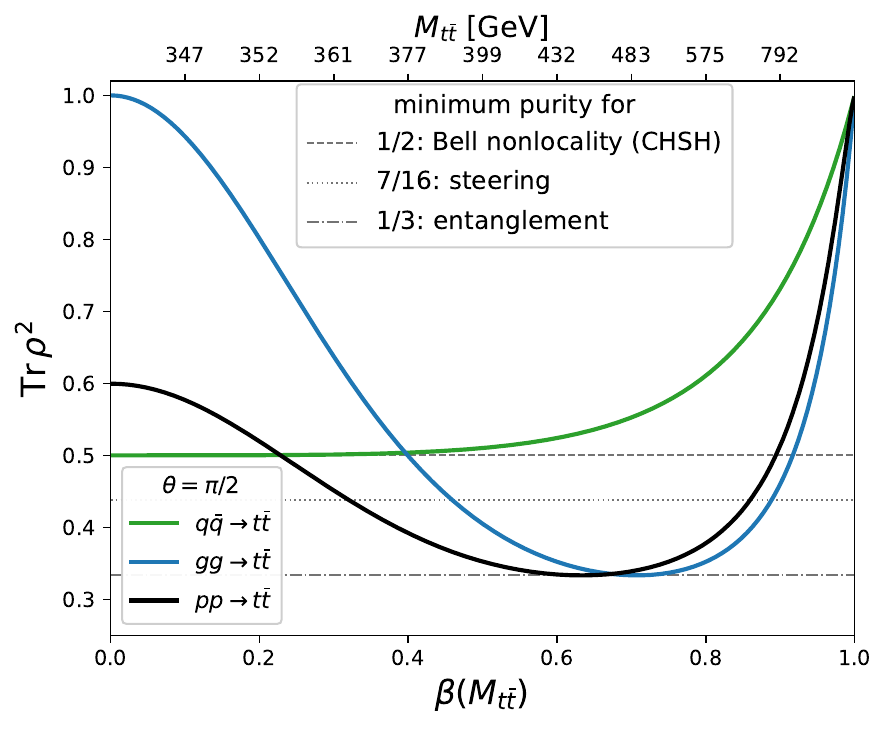}  
 \caption{Purity of the
leading-order state for $q\bar q\to t\bar t$ (top left), $gg\to t\bar t$
(top right) and $pp\to t\bar t$ (bottom left) over phase space, and at
$\theta=\pi/2$ as a function of $\beta$ for the three channels (bottom right),
at $\sqrt s=13$~TeV. The black contours and
the horizontal lines mark the minimum purities $1/3$, $7/16$ and $1/2$ required
for entanglement, steering and violation of the CHSH inequality, respectively.}
\label{fig:purity}
\end{figure}

The degree of mixing of the state is measured by its purity. For vanishing
polarizations, Eqs.~\eqref{eq:fano} and \eqref{eq:belleig} give
\begin{equation}
  \mathrm{Tr}\,\rho^2=\sum_B\lambda_B^2=\frac14\Big(1+\sum_i c_i^2\Big)
  =\frac14\Big(1+\mathrm{tr}\,C^TC\Big),
  \label{eq:purity}
\end{equation}
which ranges from $1/4$ for the maximally mixed state to $1$ for a pure state.
We show the purity of the individual and full production channels for the $t\bar{t}$ state in Fig.~\ref{fig:purity}. At $\theta=\pi/2$, Eqs.~\eqref{eq:rhoqq_bell} and
\eqref{eq:rhogg_bell} give
\begin{equation}
  \mathrm{Tr}\,\rho_{q\bar q}^2=\frac{1+a^2}{2},\qquad
  \mathrm{Tr}\,\rho_{gg}^2=\frac{(1-\beta^2)^2+\beta^4+4\beta^4(1-\beta^2)^2}
  {(1+2\beta^2-2\beta^4)^2},
  \label{eq:purity_halfpi}
\end{equation}
with $a$ as in Eq.~\eqref{eq:Cqq}. In $q\bar q$ annihilation, helicity conservation leaves only the two
initial configurations with $J_z=\pm1$, each of which produces a pure spin
state, so the $q\bar q$ state has rank two everywhere in phase space and its
purity never falls below $1/2$, and equals $1/2$ at threshold and in the forward
direction, and approaches $1$ for transverse top quarks at high $\beta$. The $gg$ state is the pure singlet at
threshold, passes at $\theta=\pi/2$ through the maximally mixed point of the
face of Fig.~\ref{fig:face}, with $\mathrm{Tr}\,\rho^2=1/3$ at $\beta^2=1/2$
($\Mtt\simeq490$~GeV), and becomes nearly maximally mixed in the forward
direction at high $\beta$. In the physical $pp$ state the $q\bar q$ admixture
lowers the purity at threshold to $\simeq0.6$ at $\theta=\pi/2$.

Although we examine the quantum correlations between the top and anti-top pair based on the leading order density matrix, the decoherence effects from high-order corrections are not expected to be appreciably large, based on the recent calculations of collinear resummation \cite{Gu:2025ijz} and the leading order radiation  \cite{Cheng:2026zfb}.

\subsection{States over phase space and fictitious states}
\label{sec:fictitious_intro}

The state measured in a region $\Pi$ of phase space is the
cross-section-weighted average
\begin{equation}
\rho_{t\bar{t}}(\Pi)=\frac{1}{\sigma_\Pi}\int_\Pi d\Omega\,\frac{d\sigma}{d\Omega}\,\rho(\Omega)
=\frac{1}{\sigma_\Pi}\int_\Pi d\Omega\,\frac{d\sigma}{d\Omega}\sum_{i,j}\rho(\Omega)_{ij}\,|i\rangle\langle j|,
\label{eq:full_qstate}
\end{equation}
where $\Omega$ collects the kinematic variables of the pair, $\sigma_\Pi$ is
the cross section in $\Pi$, and
$\ket{i}=\ket{\text{spin of}\;t}\otimes\ket{\text{spin of}\;\bar{t}}$ are the
basis states defined with respect to the chosen quantization axes. Whether
Eq.~\eqref{eq:full_qstate} is a genuine quantum state depends on these
axes~\cite{Afik:2022kwm,Han:2023fci,Cheng:2023qmz}. If the axes are fixed in
the laboratory, Eq.~\eqref{eq:full_qstate} is the density matrix of the
ensemble of pairs in $\Pi$, and all quantum-information tools apply to it.
Since $pp$ collisions are symmetric about the beam axis, however, such a
state is averaged over the azimuthal angle $\phi$ of the production plane,
and because the spin correlations of the individual pairs are referred to axes
that rotate with $\phi$, the average partially cancels them: at high $p_T$,
where each pair is close to the pure triplet $T_n$, the $\phi$-averaged
fixed-basis state is far from pure and its transverse correlations
vanish~\cite{Han:2023fci,Cheng:2023qmz}.

In the helicity basis the axes are defined event by event,
$\{|i\rangle\}=\{|i(\hat k)\rangle\}$, and Eq.~\eqref{eq:full_qstate} is then not the density matrix of any
ensemble. It is a \textit{fictitious} state~\cite{Afik:2022kwm}, whose spin
correlation matrix $\bar C_{ij}$ is the cross-section-weighted average of the
correlation matrices of the substates $\rho(\Omega)$, as defined in Eq.~\eqref{eq:Cbar} later.
Linear functions of $C$ commute with this average, so $\bar C_{ij}$ is exactly
what the decay method measures (Sec.~\ref{sec:methods}), but non-linear
quantities such as concurrence, Bell operators, discord, steerability do not,
and a quantity evaluated on $\bar C$ requires interpretation. For
entanglement and Bell nonlocality this interpretation is simple since the
separable and the Bell-local states form convex sets, an entangled or
Bell-nonlocal fictitious state certifies the existence of entangled or
Bell-nonlocal substates~\cite{Han:2023fci,Cheng:2023qmz}. For quantum discord
it fails, because the zero-discord states do not form a convex
set~\cite{Han:2024ugl}. We show in Sec.~\ref{sec:fictitious} that steering
behaves like entanglement in this respect, and that the same convexity makes
the choice of the phase-space region part of the measurement.

\begin{figure}
\centering
  \includegraphics[scale=0.45]{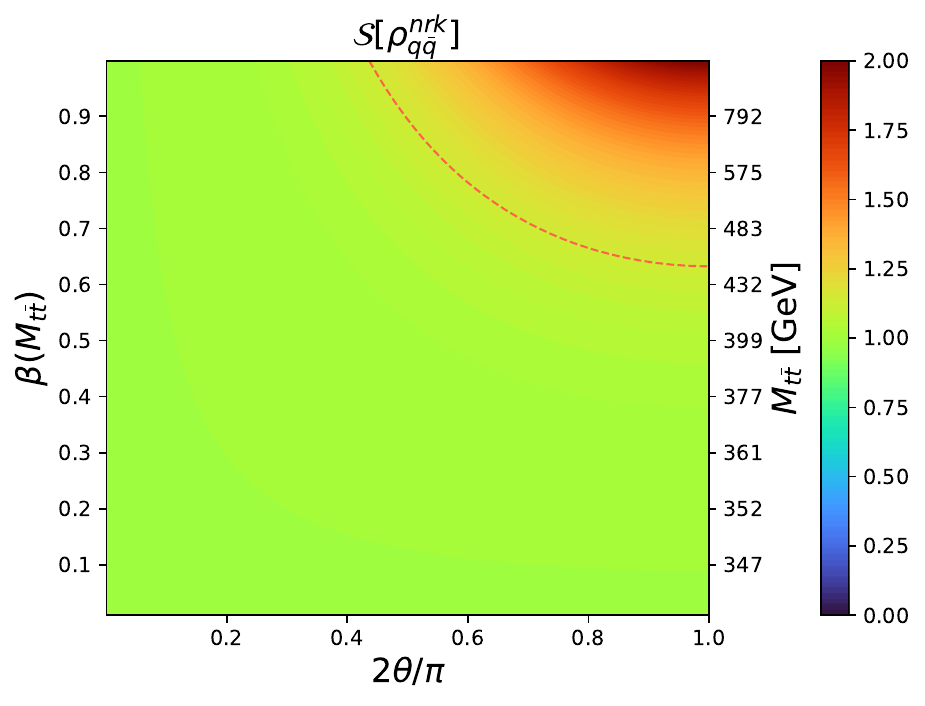}
  \includegraphics[scale=0.45]{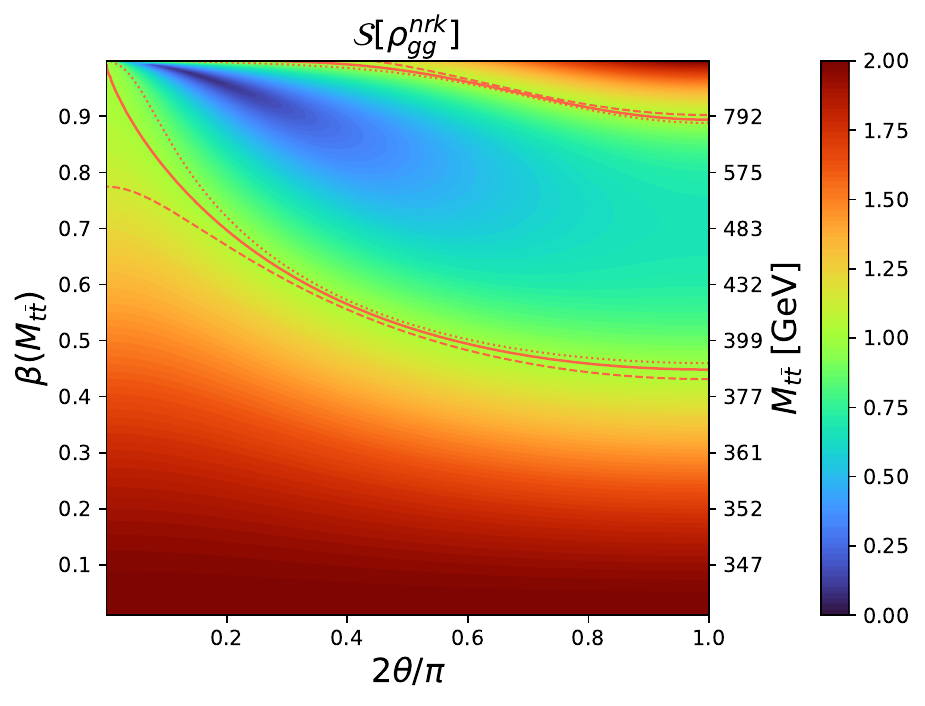}\\
  \includegraphics[scale=0.45]{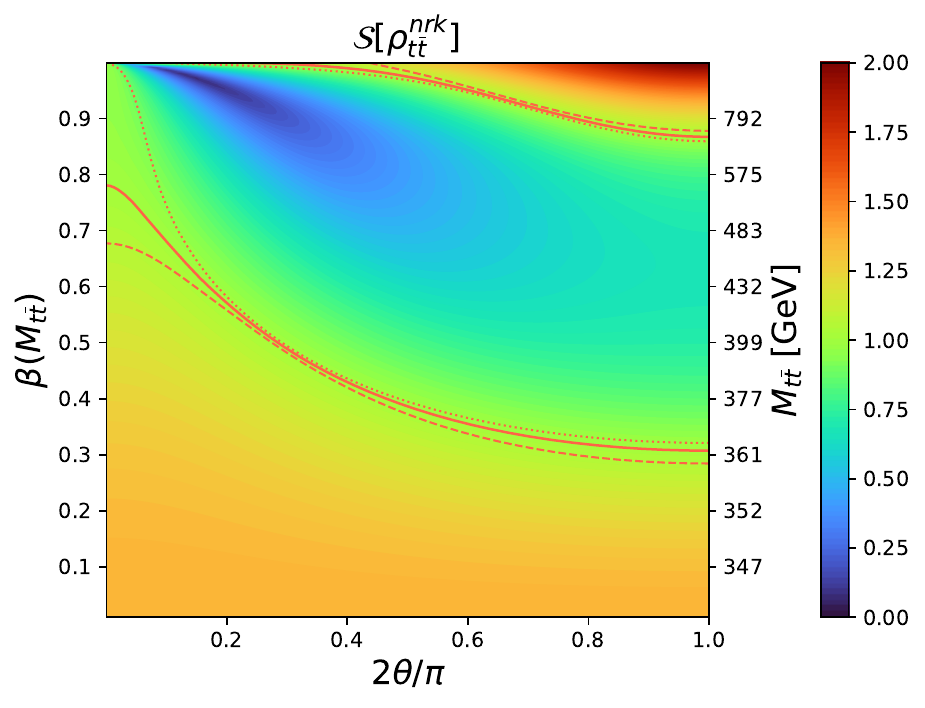}  
  \includegraphics[scale=0.45]{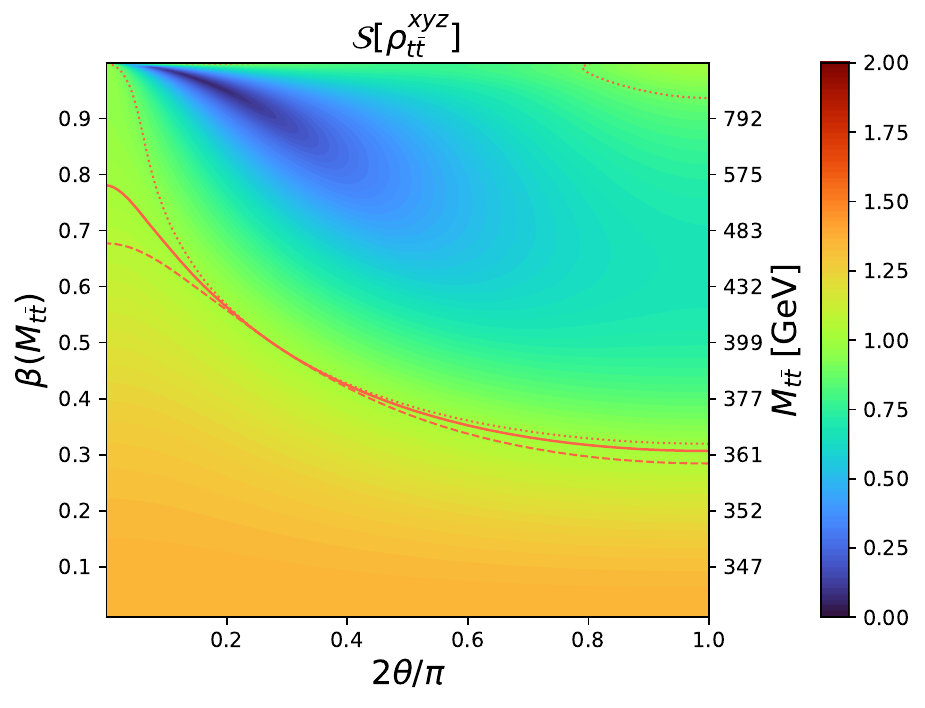}  
 \caption{The steerability integral $\Ssteer$, Eq.~\eqref{eq:SI}, for
$q\bar q\to t\bar t$ (top left), $gg\to t\bar t$ (top right) and $pp\to t\bar t$
(bottom left) in the helicity basis, and for $pp\to t\bar t$ in the beam basis, after averaging over
the azimuthal angle of the production plane (bottom right), at $\sqrt s=13$~TeV. The right axis shows the value of $M_{t\bar t}$
corresponding to $\beta$. The solid lines are the exact steerability boundary
$\Ssteer=1$, Eq.~\eqref{eq:SIcrit}; the dashed and dotted lines are the
boundaries of the closed-form sufficient and necessary conditions,
Eqs.~\eqref{eq:suff} and \eqref{eq:nec}. The $q\bar q$ state
is a rank-two mixture of two Bell states and is steerable for all $\beta>0$,
so only the sufficient-condition boundary appears in that panel.}
\label{fig:SI}
\end{figure}

\section{Steerability of $\ttbar$}
\label{sec:steerability}

\subsection{EPR steering and local hidden state models}
\label{sec:steer-def}

Consider a bipartite state $\rho_{AB}$ shared between two parties $A$ and $B$. Let $A$ perform a measurement labelled by $x$ with outcome $a$, described
by positive operators $M_{a|x}$ with $\sum_a M_{a|x}=I$. The unnormalized
states left for $B$,
\begin{equation}
\sigma_{a|x} = \mathrm{tr}_A\!\left[(M_{a|x}\otimes I)\,\rho_{AB}\right],
\qquad \sum_a \sigma_{a|x} = \rho_B ,
\label{eq:assemblage}
\end{equation}
form an \emph{assemblage}. The assemblage admits a local hidden state (LHS)
model if there exist a classical variable $\lambda$ distributed according to
$p(\lambda)$, response functions $p(a|x,\lambda)$, and a fixed ensemble of
states $\rho_\lambda$ of $B$ such that
\begin{equation}
\sigma_{a|x} = \sum_\lambda p(\lambda)\, p(a|x,\lambda)\, \rho_\lambda
\qquad \text{for all } a,\,x .
\label{eq:LHS}
\end{equation}
When such a model exists, $B$'s conditional states can be understood as a
pre-existing ensemble $\{p(\lambda),\rho_\lambda\}$ about which $A$ merely
holds classical information: $A$'s choice of measurement selects a classical
post-processing of the ensemble but does not alter the states themselves. If
no LHS model exists, $A$ can \emph{steer} $B$, and $\rho_{AB}$ is said to be
steerable from $A$ to $B$~\cite{Wiseman:2007hyt}. Steering formalizes the original
EPR argument as a one-sided task: $B$ trusts the quantum description of his
own system but makes no assumption about $A$'s devices.

Two properties follow directly from Eq.~\eqref{eq:LHS}. A separable state
$\rho_{AB}=\sum_\lambda p_\lambda\,\rho^A_\lambda\otimes\rho^B_\lambda$ always
admits an LHS model, with $p(a|x,\lambda)=\mathrm{tr}(M_{a|x}\rho^A_\lambda)$,
so steerable states are entangled. Conversely, since the statistics of any
measurement by $B$ on the states $\rho_\lambda$ are reproduced by a local
hidden variable model, Bell nonlocal states are steerable. Both inclusions are
strict for mixed states~\cite{Wiseman:2007hyt}, which is the hierarchy
quoted in the introduction. Steerability depends on the set of measurements
that $A$ is allowed to perform; throughout this work we consider projective
measurements, the class relevant for the spin measurements performed through
the decay products of the top quarks. Finally, Eq.~\eqref{eq:LHS} is not
symmetric under $A\leftrightarrow B$, and there exist states that are
steerable in one direction only. We will see that the leading-order $\ttbar$
state is symmetric in this respect.

We reiterate, however, the measurements by constructing kinematic distributions of particle decay products explore the intrinsic quantum nature, but they are qualitatively different from the experimental setting originally  proposed by John Bell \cite{Bell:1964kc}. 
It therefore should not be interpreted as to exclude a generic Local Hidden Variable Theory (LHVT) \cite{Abel:1992kz,Low:2025aqq,Abel:2025skj}, although a specific LHVT could be falsified and thus excluded by comparing with the quantum mechanical predictions  \cite{Go_2007}.

\subsection{Steering criterion for Bell-diagonal states}
\label{sec:steer-criterion}

For a two-qubit state in the Fano--Bloch form of Eq.~\eqref{eq:fano}, if $A$
measures the spin along the unit vector $\hat n$ and obtains the outcome
$\pm$, which happens with probability $p_\pm = (1\pm\hat n\cdot \vec B^{+})/2$,
the state of $B$ collapses to a qubit with Bloch vector
\begin{equation}
\vec b_{\pm}(\hat n) = \frac{\vec B^{-} \pm C^{T}\hat n}{1\pm \hat n\cdot\vec B^{+}} .
\label{eq:condbloch}
\end{equation}
For the Bell-diagonal states relevant at leading order, $\vec B^{\pm}=0$ and
$\vec b_\pm(\hat n) = \pm C^{T}\hat n$. As $\hat n$ ranges over the unit
sphere these vectors sweep out the \emph{quantum steering
ellipsoid}~\cite{Jevtic:2014icx}, centered at the origin of $B$'s Bloch ball,
whose semi-axes are the singular values $\mu_1\ge\mu_2\ge\mu_3\ge 0$ of the
spin correlation matrix $C$.

Jevtic \emph{et al.}~\cite{Milne:2014nsw} constructed explicit LHS models for
Bell-diagonal states and showed that the state is unsteerable with respect to
projective measurements whenever the \emph{steerability integral}
\begin{equation}
\Ssteer \;\equiv\; \frac{1}{2\pi}\int d\hat n\, \sqrt{\hat n^{T} C C^{T}\, \hat n}
\;=\; 2\,\big\langle\, |C^{T}\hat n|\, \big\rangle
\label{eq:SI}
\end{equation}
satisfies $\Ssteer\le 1$, where $\langle\cdot\rangle_{\hat n}$ denotes the
uniform average over the sphere. The converse, that $\Ssteer>1$ guarantees
steerability, was subsequently proven in Ref.~\cite{Nguyen:2016uga}, so that
\begin{equation}
\Ssteer > 1 \quad\Longleftrightarrow\quad \rho_{AB}\ \text{is steerable}
\label{eq:SIcrit}
\end{equation}
for all Bell-diagonal two-qubit states and projective
measurements.\footnote{For states with non-vanishing polarizations the sufficient conditions for unsteerability are given in
Ref.~\cite{Bowles:2016vky} and a complete solution in terms of the ``critical
radius'' in Ref.~\cite{Nguyen:2016uga}.}
Equation~\eqref{eq:SI} has a simple interpretation: $\Ssteer$ is twice the
length of $B$'s conditional Bloch vector, averaged uniformly over $A$'s
measurement directions. The state is steerable when $A$ can displace $B$'s
state, on average, by more than half the radius of the Bloch ball. The
integral lies in the range $0\le\Ssteer\le2$, with $\Ssteer=2$ for the four
Bell states.

Three properties of $\Ssteer$ are important for the collider application.
\begin{itemize}
\item $\Ssteer$ depends on $C$ only through its singular values. It is
  therefore invariant under independent rotations of the quantization axes of
  $t$ and $\bar t$ and, for a quantum state at a given point in phase space, it
  takes the same value in the helicity and beam bases. The difference, thus, arises due to the averaging over phase space, and it becomes crucial to interpret the state as a fictitious state.
\item $\Ssteer$ is a convex function of $C$. For fixed $\hat n$ the map
  $C\mapsto|C^{T}\hat n|$ is a seminorm, and the average of convex functions
  is convex. This property underlies the discussion of fictitious states in
  Sec.~\ref{sec:fictitious}.
\item Since $C^{T}C$ and $CC^{T}$ have the same spectrum, $\Ssteer$ is
  symmetric under $A\leftrightarrow B$: for Bell-diagonal states, $t$ can
  steer $\bar t$ if and only if $\bar t$ can steer $t$. One-way steering
  requires $\vec B^{\pm}\neq0$ and can only arise in $\ttbar$ production
  beyond leading order or from new physics, in close analogy with the
  asymmetry of quantum discord as a probe of CP violation~\cite{Afik:2022dgh}.
\end{itemize}

It is useful to compare Eq.~\eqref{eq:SIcrit} with the criteria for the
neighboring levels of the hierarchy. A Bell-diagonal state is entangled if and
only if the largest eigenvalue of $\rho$ exceeds $1/2$, which in terms of the
eigenvalues of $C$ reproduces the conditions used in
Refs.~\cite{Aguilar-Saavedra:2022uye,Han:2023fci}. The CHSH inequality can be
violated if and only if $\mu_1^2+\mu_2^2>1$~\cite{Horodecki:1995nsk}. Steering can
also be certified by linear steering inequalities with a finite number $m$ of
measurement settings~\cite{Cavalcanti:2009wia}; for $m=2$ and $m=3$ orthogonal
settings these are optimally violated when $\mu_1+\mu_2>\sqrt2$ and
$\mu_1+\mu_2+\mu_3>\sqrt3$, respectively. Each of these conditions is
sufficient for $\Ssteer>1$ but, as we show next, Eq.~\eqref{eq:SIcrit} is
strictly more powerful.

\subsection{Closed-form steering criteria}
\label{sec:steer-bounds}

The integral in Eq.~\eqref{eq:SI} has no elementary closed form for general
$\mu_i$, but it is tightly bracketed by two elementary symmetric functions of
the singular values,
\begin{equation}
\frac{2}{3}\,\big(\mu_1+\mu_2+\mu_3\big)\;\le\;\Ssteer\;\le\;
2\sqrt{\frac{\mu_1^2+\mu_2^2+\mu_3^2}{3}} .
\label{eq:bounds}
\end{equation}
The upper bound follows from the concavity of the square root together with
$\langle \hat n_i^2\rangle_{\hat n}=1/3$. For the lower bound, note that for
fixed $\hat n$ the integrand $(\sum_i\mu_i^2 \hat n_i^2)^{1/2}$ is a convex
function of $(\mu_1,\mu_2,\mu_3)$, that the integral is invariant under
permutations of the $\mu_i$ because the measure on the sphere is, and that a
convex permutation-symmetric function $f$ obeys
$f(\mu)\ge f(\bar\mu,\bar\mu,\bar\mu)$ with $\bar\mu$ the mean of the $\mu_i$.
Since $\Ssteer(\bar\mu,\bar\mu,\bar\mu)=2\bar\mu$, the lower bound follows.
Both bounds are saturated for $\mu_1=\mu_2=\mu_3$, i.e. for Werner-like
states, and they lie within a few percent of $\Ssteer$ for the $\ttbar$ states
encountered below. Eq.~\eqref{eq:bounds} immediately yields a sufficient
and a necessary condition for EPR steerability,
\begin{align}
\sum_i \mu_i \;&>\; \frac32 \qquad\;\;\Longrightarrow\quad \text{steerable},
\label{eq:suff}\\
\mathrm{tr}\big(C^{T}C\big) = \sum_i \mu_i^2 \;&>\; \frac34
\qquad\Longleftarrow\quad \text{steerable},
\label{eq:nec}
\end{align}
both of which are exact for Werner states. These are the closed-form criteria
referred to in the introduction, and they are the curves shown in
Fig.~\ref{fig:SI}.

Two comments are in order. First, the sufficient condition~\eqref{eq:suff} is
strictly stronger than the three-setting linear steering inequality
$\sum_i\mu_i>\sqrt3\simeq1.73$: a state with $3/2<\sum_i\mu_i\le\sqrt3$ is
steerable, but its steerability cannot be revealed by three orthogonal
measurement settings and requires the full set of projective measurements that
enters Eq.~\eqref{eq:SI}. The Werner states with $1/2<p\le1/\sqrt3$ are an
example. Second, the necessary condition~\eqref{eq:nec} involves only the
basis-independent quantity $\mathrm{tr}(C^{T}C)=\sum_{ij}C_{ij}^2$, which is a
natural target for a first experimental test: a measured value
$\mathrm{tr}(C^{T}C)\le3/4$ excludes steerability without any need to extract
the singular values. For the leading-order $\ttbar$ state in the helicity
basis only $C_{kk}$, $C_{rr}$, $C_{nn}$ and $C_{kr}=C_{rk}$ are non-zero, so
that $\mu_n=|C_{nn}|$ and the remaining two singular values are the absolute
values of the eigenvalues of the $2\times2$ block in the $(k,r)$ plane.

\subsection{Steerability of the leading-order $\ttbar$ state}
\label{sec:steer-ttbar}

The structure of the steerability of $\ttbar$ over phase space, shown in
Fig.~\ref{fig:SI}, can be understood analytically in the transverse
configuration $\theta=\pi/2$, where the state lies on the face of the Bell
tetrahedron shown in Fig.~\ref{fig:face} and the spin correlation matrices
take the closed forms of Eqs.~\eqref{eq:Cqq} and \eqref{eq:Cgg}. 

For $q\bar q$ annihilation the correlation along the beam direction $r$ is
complete for all $\beta$, reflecting the spin-triplet, $J_z=\pm1$
configuration of the initial state, while the transverse correlations grow
with $\beta$. The singular values are $\mu=(1,a,a)$ and the integral in
Eq.~\eqref{eq:SI} can be done in closed form,
\begin{equation}
\Ssteer_{q\bar q}\big(\beta,\tfrac{\pi}{2}\big) =
1 + \frac{a^2}{\sqrt{1-a^2}}\,\ln\!\frac{1+\sqrt{1-a^2}}{a} .
\label{eq:Sqq}
\end{equation}
It increases monotonically from $\Ssteer=1$ at threshold, where the state is
the separable mixture correlated along the beam, to $\Ssteer=2$ for
$\beta\to1$, where the state becomes the pure triplet. The $q\bar q$ state at
$\theta=\pi/2$ is a rank-two mixture of two Bell states with eigenvalues
$(1\pm a)/2$, for which entanglement, steerability and Bell nonlocality set in
simultaneously at $\beta>0$: the concurrence is $\mathcal{C}=a$, $\Ssteer>1$,
and $\mu_1^2+\mu_2^2=1+a^2>1$. The sufficient condition~\eqref{eq:suff} is met
for $a>1/4$, i.e.\ $\beta>\sqrt{2/5}$ or $\Mtt>445$~GeV.

Gluon fusion is richer. The state interpolates between the singlet at
threshold and the triplet at high $p_T$, passing through a strongly mixed,
separable state at $\beta^2=1/2$ where all three singular values equal $1/3$
and $\Ssteer=2/3$. Evaluating Eq.~\eqref{eq:SI} numerically with
Eq.~\eqref{eq:Cgg}, the $gg$ state at $\theta=\pi/2$ is steerable for
\begin{equation}
\beta<0.448\ \ (\Mtt<386~\mathrm{GeV}) \qquad\text{or}\qquad
\beta>0.894\ \ (\Mtt>770~\mathrm{GeV}).
\label{eq:ggwindow}
\end{equation}
For comparison, the same state is entangled for
$\Mtt<410$~GeV or $\Mtt>637$~GeV and violates the CHSH inequality for
$\Mtt<371$~GeV or $\Mtt>942$~GeV. The three windows are nested as required by
the hierarchy, and they make the two distinctive features of steering at the
LHC quantitative: between $770$ and $942$~GeV the $gg$ state at $\theta=\pi/2$
is steerable but not Bell nonlocal, while between $386$ and $410$~GeV and
between $637$ and $770$~GeV it is entangled but not steerable. Away from
$\theta=\pi/2$ the correlations weaken and the steerable windows shrink, as
seen in Fig.~\ref{fig:SI}.

The physical $pp\to\ttbar$ state is the luminosity-weighted mixture of
Eq.~\eqref{eq:rhoThetaMtt}. Since $\Ssteer$ is convex in $C$, mixing can only
lower the steerability integral below the weighted average of its $q\bar q$
and $gg$ values. Near threshold the $q\bar q$ admixture, with
$\Ssteer_{q\bar q}\simeq1$, dilutes the strongly steerable singlet and pushes
the low-mass boundary of Eq.~\eqref{eq:ggwindow} to lower $\Mtt$; in the
boosted regime both channels approach the same triplet state and the dilution
is mild. This is why in Fig.~\ref{fig:SI} (bottom left) the steerable region at
threshold is confined to $\Mtt\lesssim360$~GeV at central angles, reaching $\Mtt\simeq400$--$450$~GeV only in the forward direction, while the boosted region extends
over a wide range of $\theta$ once $\Mtt\gtrsim800$~GeV. The bottom right panel of Fig.~\ref{fig:SI} shows
the beam basis. Since $\Ssteer$ is basis independent for a quantum state at a
point in phase space, the difference between the two bottom panels arises
entirely from the average over the azimuthal angle of the production plane
that is implicit in a fixed-basis
state~\cite{Han:2023fci,Cheng:2023qmz}, which washes out the
transverse correlations at high $p_T$.

\subsection{Fictitious states}
\label{sec:fictitious}

As discussed in Sec.~\ref{sec:quantum_states}, the state reconstructed in a region
$\Pi$ of phase space in the helicity basis is a fictitious state, specified by
the cross-section-weighted average of the spin correlation matrices of the
substates,
\begin{equation}
\bar C_{ij} = \frac{1}{\sigma_\Pi}\int_\Pi d\Omega\,\frac{d\sigma}{d\Omega}\,C_{ij}(\Omega)
= \sum_a w_a\, C^{(a)}_{ij},
\qquad w_a\ge0,\quad \sum_a w_a = 1,
\label{eq:Cbar}
\end{equation}
where in the second form the region has been discretized into substates $a$
with weights $w_a$. Each $C^{(a)}$ is evaluated in its own helicity frame, so
$\bar C$ does not correspond to a physical density matrix, and the question is
what a measurement of $\Ssteer(\bar C)>1$ tells us about the genuine quantum
states $C^{(a)}$.

Writing
$\vec v_a = (C^{(a)})^{T}\hat n$ and using the triangle inequality,
\begin{equation}
\Ssteer(\bar C) = \frac{1}{2\pi}\int d\hat n\,\Big|\sum_a w_a \vec v_a\Big|
\;\le\; \sum_a w_a\, \frac{1}{2\pi}\int d\hat n\,|\vec v_a|
= \sum_a w_a\, \Ssteer\big(C^{(a)}\big) .
\label{eq:convex}
\end{equation}
If every substate is unsteerable, $\Ssteer(C^{(a)})\le1$ for all $a$, then
$\Ssteer(\bar C)\le1$ and the fictitious state is unsteerable as well.
Equivalently,
\begin{equation}
\Ssteer(\bar C) > 1 \quad\Longrightarrow\quad
\text{there exists a substate, } a:\ \Ssteer\big(C^{(a)}\big)>1 ,
\label{eq:witness}
\end{equation}
so that a steerable fictitious state certifies the existence of at least one
steerable genuine quantum state in the region $\Pi$, in the same way as
entanglement and Bell nonlocality of fictitious states certify entangled and
Bell-nonlocal substates~\cite{Han:2023fci,Cheng:2023qmz}. This is
the collider analog of the statement that the set of unsteerable states is
convex~\cite{Wiseman:2007hyt}, and it contrasts with quantum discord, for which
the set of zero-discord states is not convex and additional information about
the substates is needed to interpret a measurement~\cite{Han:2024ugl}.

Eq.~\eqref{eq:convex} has two practical corollaries. First, averaging
over phase space can only reduce the steerability integral below the average
of its pointwise values, so an inclusive selection can fail to be steerable
even when every one of its substates is: the converse of
Eq.~\eqref{eq:witness} does not hold. Steerability is therefore best measured
in compact regions of phase space where the substates are both strongly
steerable and similar to one another, and the choice of signal region is part
of the measurement. We return to this point in Sec.~\ref{sec:regions}, where
we find that the threshold region used in previous entanglement and discord
studies lies at the steerability boundary, while the boosted region is
robustly steerable. Second, since $\Ssteer$ is convex, statistical
fluctuations of the measured $\bar C$ bias the estimator $\Ssteer(\bar C)$
upwards by an amount of order $\delta C^2$. For the uncertainties achieved
below this bias is well below the statistical uncertainty, but it should be
kept in mind whenever $\Ssteer$ is close to one.

\section{Collider analysis}
\label{sec:collider_analysis}

We now turn to the measurement of the steerability of the $\ttbar$ state at
the LHC. The analysis parallels the studies of
entanglement~\cite{Han:2023fci} and quantum discord~\cite{Han:2024ugl}, to which we refer for details, and we summarize here the
aspects specific to steering. We consider both the dileptonic and the
semileptonic decay channels,
\begin{equation}
pp\to\ttbar\to b\bar b\,\ell^+\nu\,\ell^-\bar\nu, \qquad\qquad
pp\to\ttbar\to b\bar b\,\ell^\pm\nu\,q\bar q',
\label{eq:channels}
\end{equation}
with $\ell=e,\mu$. The dileptonic channel offers the maximal spin analyzing
power of the two charged leptons, $|\kappa_\ell|=1$, at the price of a small
branching fraction and a two-neutrino final state. The semileptonic channel
has a branching fraction about six times larger and a single neutrino, which
makes the kinematic reconstruction considerably more efficient, at the price
of a reduced spin analyzing power on the hadronic side. Using the optimal
hadronic direction of Ref.~\cite{Tweedie:2014yda}, a weighted combination of the
two light-quark jets, the hadronically decaying top is analyzed with
$|\kappa_{\rm opt}|=0.64$. The two effects together make the semileptonic
channel about $60\%$ more sensitive than the dileptonic one for a fixed
luminosity, before reconstruction efficiencies are taken into
account~\cite{Han:2023fci}.

\subsection{Event simulation and reconstruction}
\label{sec:simulation}

Events are generated at leading order for $pp\to\ttbar$ at $\sqrt{s}=13$~TeV
with \texttt{MadGraph5}~\cite{Alwall:2011uj} and decayed with
\texttt{MadSpin}~\cite{Artoisenet:2012st}. We use the \texttt{NNPDF4.0} NNLO PDF set with
$\alpha_{s}(m_{Z})=0.118$~\cite{NNPDF:2021njg} included in the
\texttt{LHAPDF6} package~\cite{Buckley:2014ana}. In evaluating the PDFs, we
set the factorization scale at $Q=M_{t\bar{t}}$. Parton
showering and hadronization are performed with
\texttt{Pythia~8}~\cite{Bierlich:2022pfr} and the detector response is
simulated with \texttt{Delphes~3}~\cite{deFavereau:2013fsa}. 

For the dileptonic channel we follow Ref.~\cite{Han:2024ugl}: we require at
least two jets with $p_T>25$~GeV and $|\eta|<2.5$, at least one of which is
$b$-tagged, and exactly two isolated opposite-sign leptons with $p_T>25$~GeV
and $|\eta|<2.5$. The two neutrino momenta are obtained by solving the
kinematic constraints from the $W$ and top masses~\cite{Sonnenschein:2006ud},
smearing the inputs within the detector resolution and weighting the solutions
by the truth-level $m_{b\ell}$ distribution, following the strategy of the CMS
and ATLAS spin correlation and entanglement
measurements~\cite{CMS:2019nrx,ATLAS:2023fsd}. For the semileptonic channel we
follow Ref.~\cite{Han:2023fci}: we require jets and a lepton with
$p_T>25$~GeV and $|\eta|<2.5$ and $E_T^{\rm miss}>30$~GeV, two $b$-tagged jets
(or the hardest untagged jet as the second $b$-jet candidate), and we
reconstruct the neutrino from the missing transverse momentum and the $W$ and
top on-shell conditions with a pseudo-top algorithm. This resolved
reconstruction is used in the threshold region. In the boosted region the
decay products of the hadronically decaying top quark merge, and we use
instead the boosted-topology reconstruction of Ref.~\cite{Han:2023fci}: the
event is reclustered into large-radius jets ($R=1.5$, $p_T>300$~GeV), each of
which is matched to the three highest-$p_T$ small-radius subjets ($R=0.2$)
that carry more than $90\%$ of its transverse momentum; the hadronic top is
the four-vector sum of the three subjets, one of which is required to be
$b$-tagged, and its mass is required to lie in the window
$150$--$225$~GeV. This selection recovers an efficiency of several percent up
to $\Mtt\simeq2$~TeV, where the resolved reconstruction fails by more than an
order of magnitude~\cite{Han:2023fci}. The average reconstruction efficiencies
in the signal regions defined below are listed in Table~\ref{tab:eff}.

\begin{table}[t]
\centering
\begin{tabular}{l cc @{\hspace{2.5em}} cc}
 & \multicolumn{2}{c}{Dileptonic} & \multicolumn{2}{c}{Semileptonic} \\
\midrule
 & Threshold & Boosted & Threshold & Boosted \\
\midrule
$\langle\epsilon_{\rm rec}\rangle$ & 0.10 & 0.08 & 0.13 & 0.07 \\
\bottomrule
\end{tabular}
\caption{Average reconstruction efficiency in the signal regions of
Sec.~\ref{sec:regions} at $\sqrt s=13$~TeV. The semileptonic values use the
resolved reconstruction in the threshold region and the boosted-topology
reconstruction in the boosted region.}
\label{tab:eff}
\end{table}

\subsection{Reconstruction of the spin correlation matrix}
\label{sec:methods}

The spins are measured with respect to a set of quantization axes. The
helicity basis $\{\hat k,\hat r,\hat n\}$ is defined in the $\ttbar$
center-of-mass frame with $\hat k$ along the top-quark momentum. Since $\Ssteer$ is basis independent event
by event, the choice of basis matters only through the averaging over phase
space discussed in Sec.~\ref{sec:fictitious}. We present results in the
helicity basis, in which the averaged correlations are
largest~\cite{Han:2023fci,Cheng:2023qmz}. 

The steerability integral is a function of the spin correlation matrix only,
so the task is to reconstruct $\bar C_{ij}$ in a given region of phase space.
We use the two complementary methods of
Refs.~\cite{Han:2024ugl,Cheng:2024rxi}.

In the \emph{decay method} the spins are inferred from the angular
distributions of the decay products. For a decay product $a$ of the top and
$b$ of the anti-top, with momenta evaluated in the rest frames of their
parents,
\begin{equation}
\frac{1}{\sigma}\frac{d^2\sigma}{d\cos\theta^a_i\,d\cos\theta^b_j}
= \frac14\Big(1+\kappa_a\kappa_b\,C_{ij}\cos\theta^a_i\cos\theta^b_j\Big),
\label{eq:decaydist}
\end{equation}
where $\theta^a_i$ is the angle between the momentum of $a$ and the axis
$\hat i$, and $\kappa_{a,b}$ are the spin analyzing
powers~\cite{Brandenburg:2002xr}: $\kappa_\ell=1$ for the charged lepton and
$\kappa_{\rm opt}=0.64$ for the optimal hadronic direction. Each $C_{ij}$ is
extracted from the asymmetry
\begin{equation}
C_{ij} = \frac{4}{\kappa_a\kappa_b}\,
\frac{N(\cos\theta^a_i\cos\theta^b_j>0)-N(\cos\theta^a_i\cos\theta^b_j<0)}
{N(\cos\theta^a_i\cos\theta^b_j>0)+N(\cos\theta^a_i\cos\theta^b_j<0)},
\label{eq:asym}
\end{equation}
which is robust against systematic effects. The detector-level distributions
are corrected to truth level by iterative Bayesian unfolding, implemented in
\texttt{RooUnfold}~\cite{Adye:2011gm}. The uncertainties on the corrected
coefficients are obtained from pseudo-experiments and include the statistical
uncertainty of the sample and the systematic uncertainty of the correction. They are propagated to $\Ssteer$
by evaluating Eq.~\eqref{eq:SI} on each pseudo-experiment. This approach
follows the strategy of the ATLAS and CMS entanglement
measurements~\cite{ATLAS:2023fsd,CMS:2024pts}; see also Ref.~\cite{ATLAS:2026bre}
for a discussion of the general principles of such measurements.

In the \emph{kinematic method} the spin correlation matrix of each event is
computed from the reconstructed $(\theta,\Mtt)$ with the leading-order
analytic templates, Eq.~\eqref{eq:rhoThetaMtt}, weighted by the parton
luminosities, and $\bar C_{ij}$ is the average over the events in the region;
such analytic spin density matrices can also be generated automatically for
general processes~\cite{Durupt:2025wuk}. No unfolding is applied. In the boosted region the kinematic variables are
only mildly affected by detector effects, whereas in a narrow selection the
migration in $(\theta,\Mtt)$ shifts the result (Sec.~\ref{sec:results}); the
quoted uncertainty is purely statistical, given by the standard error of the
event-by-event $C_{ij}$ in the sample. Since
$\Ssteer$ is a non-linear function of $C$, we evaluate it on the averaged
matrix, $\Ssteer(\bar C)$, rather than averaging the event-by-event values.
By Eq.~\eqref{eq:convex} the latter is always larger; only the former is the
quantity measured by the decay method and the one that admits the
interpretation of Eq.~\eqref{eq:witness}.

\subsection{Signal regions}
\label{sec:regions}

The maps of Fig.~\ref{fig:SI} identify two steerable regions of phase space,
near threshold and at high $p_T$, separated by a wide unsteerable band at
intermediate $\Mtt$. We define two signal regions in the $(\theta,\Mtt)$
plane:
\begin{itemize}
\item \textbf{Threshold region:}
  \begin{align*}
  \Mtt &\le 360~\mathrm{GeV},\\
  \Mtt &\le 380~\mathrm{GeV}\quad\text{and}\quad |\cos\theta|\ge 0.80,\\
  \Mtt &\le 400~\mathrm{GeV}\quad\text{and}\quad |\cos\theta|\ge 0.98,
  \end{align*}
\item \textbf{Boosted region:}
  \begin{align*}
  \Mtt &\ge 800~\mathrm{GeV}\quad\text{and}\quad |\cos\theta|\le 0.59,\\
  \Mtt &\ge 1100~\mathrm{GeV}\quad\text{and}\quad |\cos\theta|\le 0.71,
  \end{align*}
\end{itemize}
where a region is the union of the listed selections, and the same regions
are used for the semileptonic and dileptonic channels. The boosted region is
that of Ref.~\cite{Han:2024ugl}. The threshold region is new: it is the result
of the optimization described below, and it is considerably tighter than the
inclusive threshold region used for entanglement and
discord~\cite{Aguilar-Saavedra:2022uye,Han:2023fci,Han:2024ugl}.

Two considerations enter the choice. The significance with which steerability
can be established is $(\Ssteer-1)/\delta\Ssteer$, so one wants to maximize
the excess of $\Ssteer$ over unity while retaining enough events to keep
$\delta\Ssteer$ small; and, by Eq.~\eqref{eq:convex}, the region should
contain substates that are as similar as possible. In the boosted region both
requirements are easily met: the state approaches the pure triplet, $\Ssteer$
rises monotonically with $\Mtt$ and $\theta$, and the falling cross section is
the only limitation. Near threshold the situation is more delicate. The
steerable window at $\theta=\pi/2$ closes already at $\Mtt\simeq386$~GeV for
$gg$, Eq.~\eqref{eq:ggwindow}, and at a lower value once the $q\bar q$
admixture is included, whereas the threshold regions used for entanglement and
discord extend to $400$--$600$~GeV. The average over the region therefore
includes a large fraction of unsteerable substates, and the resulting
fictitious state sits at the boundary $\Ssteer\simeq1$
(Sec.~\ref{sec:results}). The threshold region defined above is the result of
maximizing the expected significance $(\Ssteer-1)/\delta\Ssteer$, with
$\delta\Ssteer$ obtained by propagating the statistical uncertainty of the
decay method at $139~\mathrm{fb}^{-1}$ (including the reconstruction efficiency
and a factor of two for the unfolding) through Eq.~\eqref{eq:SI}, over regions
of the form $\Mtt\le M_0$ supplemented by two forward wedges, and, as a
cross-check, over the superlevel sets of the pointwise $\Ssteer$ of
Fig.~\ref{fig:SI}.\footnote{Throughout this work, we adopt the leading-order top-quark production. We have not taken into account the possible non-perturbative contributions near the threshold  \cite{Fuks:2021xje,Fucilla:2026mkg}. In particular, in this region it was recently observed that the $t\bar{t}$ cross section is enhanced due to the presence of a pseudoscalar quasi-bound state \cite{ATLAS:2025kvb,CMS:2025kzt}, which could be interpreted as a toponium state~\cite{Sumino:1992ai}. However, our results on the quantum correlation  $C_{ij}$ should remain unaffected.} The two families give the same optimum within $5\%$, with a
broad maximum around the selection adopted here, which follows the
$\Ssteer=1$ contour at a distance that balances the gain in $\Ssteer$ against
the loss of events: it retains about $20\%$ of the events of the inclusive
region but raises the steerability integral of the fictitious state from
$\Ssteer\simeq0.94$ to $\Ssteer\simeq1.19$. The selection is tighter than the
$\Mtt<380$~GeV cut used by ATLAS to observe
entanglement~\cite{ATLAS:2023fsd}, which gives $\Ssteer\simeq1.06$.

\subsection{Results at the LHC and HL-LHC}
\label{sec:results}

\begin{figure}[t]
\centering
\includegraphics[width=1\textwidth]{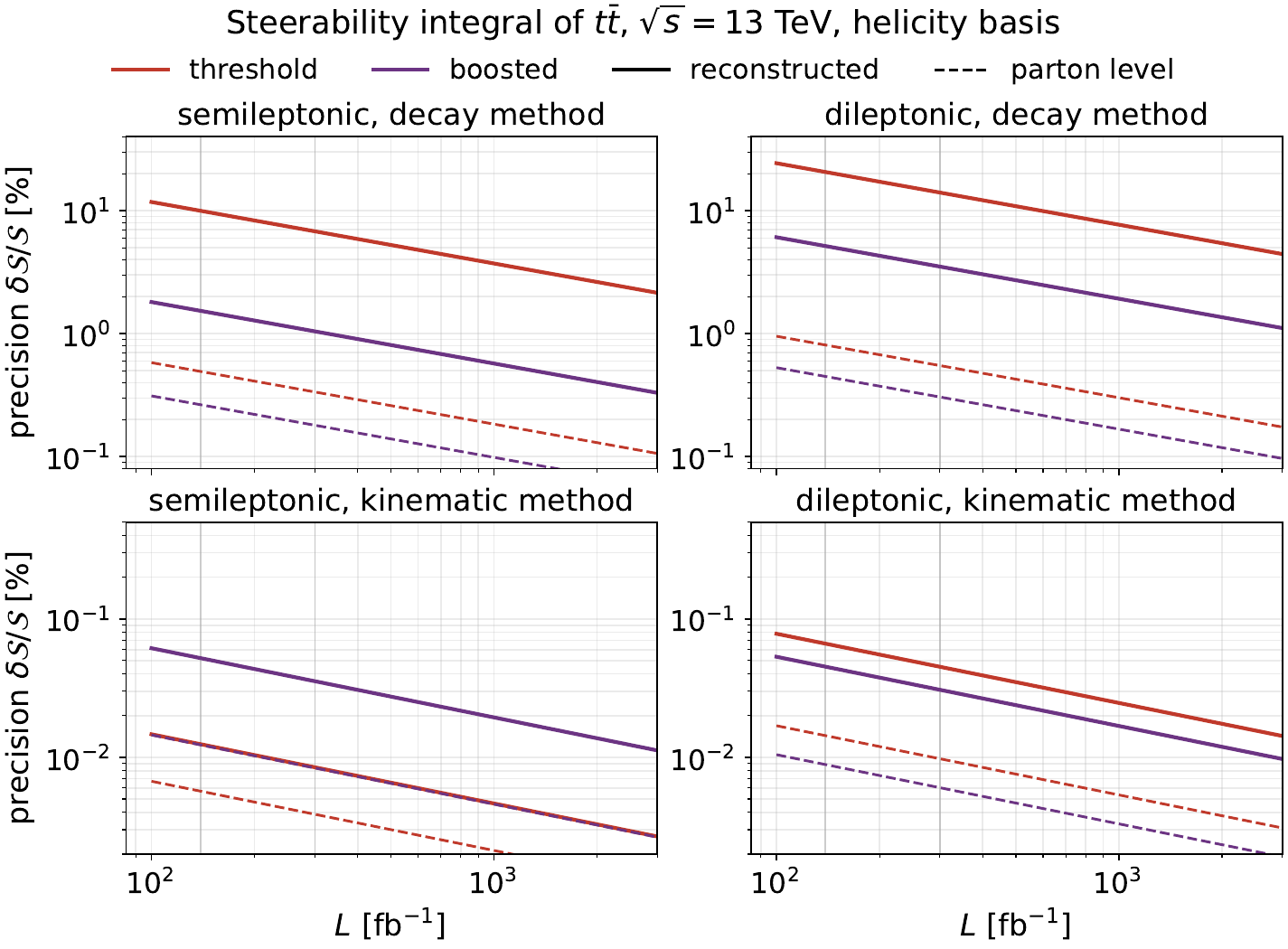}
\caption{Expected precision on the steerability integral as a function
of the integrated luminosity, at parton level (dashed) and after detector
simulation and reconstruction (solid), in the threshold (red) and boosted
(purple) regions, for both channels and both methods. Fig.~\ref{fig:signif}
shows the corresponding significances.}
\label{fig:proj}
\end{figure}

\begin{figure}[t]
\centering
\includegraphics[width=0.75\textwidth]{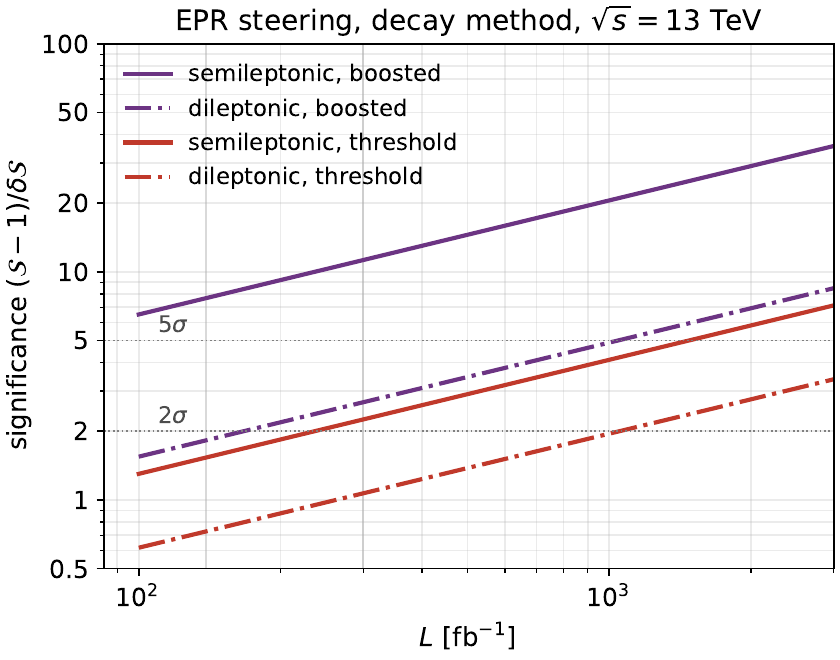}
\caption{Expected
significance for establishing steerability with the decay method in the
boosted (purple) and optimized threshold (red) regions; the horizontal lines
indicate $2\sigma$ and $5\sigma$. All results are in the helicity basis at
$\sqrt s=13$~TeV.}
\label{fig:signif}
\end{figure}

Tables~\ref{tab:SI_semilep} and~\ref{tab:SI} collect the steerability integral
in the helicity basis at parton level and after detector simulation and
reconstruction, for $\mathcal{L}=139~\mathrm{fb}^{-1}$, in the semileptonic
and dileptonic channels. We quantify the results in two ways. The
significance for establishing steerability is defined, in analogy with
Refs.~\cite{Han:2023fci,Han:2024ugl}, relative to the null hypothesis of an
unsteerable state,
\begin{equation}
\text{significance} = \frac{\Ssteer - 1}{\delta\Ssteer},
\label{eq:significance}
\end{equation}
and the precision of the measurement of the steerability integral is
$\delta\Ssteer/\Ssteer$. We
estimate both quantities to higher integrated luminosities with
$\delta\Ssteer(\mathcal{L})=\delta\Ssteer(139~\mathrm{fb}^{-1})\sqrt{139~\mathrm{fb}^{-1}/\mathcal{L}}$.
The projections for the LHC Run~3 ($300~\mathrm{fb}^{-1}$) and the HL-LHC
($3~\mathrm{ab}^{-1}$) are collected in Table~\ref{tab:proj} and shown as a
function of the luminosity in Figs.~\ref{fig:proj} and \ref{fig:signif}.

\begin{table}[t]
\centering
\begin{tabular}{l cc @{\hspace{2.5em}} cc}
& \multicolumn{2}{c}{Decay method} & \multicolumn{2}{c}{Kinematic method}\\
\midrule
& Threshold & Boosted & Threshold & Boosted\\
\midrule
Parton        & $1.1814\pm0.0058$ & $1.1335\pm0.0030$ & $1.1863\pm0.0001$ & $1.1397\pm0.0001$ \\
Reconstructed & $1.18\pm0.12$     & $1.133\pm0.017$   & $1.2705\pm0.0002$ & $1.1334\pm0.0006$ \\
\bottomrule
\end{tabular}
\caption{Steerability integral $\Ssteer$ evaluated with the decay method and
the kinematic method in the helicity basis in the semileptonic channel, at
parton level (top row) and after detector simulation, reconstruction and
unfolding (bottom row). $\Ssteer>1$ certifies EPR steerability.
$\sqrt{s}=13$~TeV, $\mathcal{L}=139$~fb$^{-1}$.}
\label{tab:SI_semilep}
\end{table}

\begin{table}[t]
\centering
\begin{tabular}{l cc @{\hspace{2.5em}} cc}
& \multicolumn{2}{c}{Decay method} & \multicolumn{2}{c}{Kinematic method}\\
\midrule
& Threshold & Boosted & Threshold & Boosted\\
\midrule
Parton        & $1.1615\pm0.0094$ & $1.113\pm0.005$ & $1.1860\pm0.0002$ & $1.1292\pm0.0001$ \\
Reconstructed & $1.18\pm0.24$     & $1.104\pm0.057$ & $1.1647\pm0.0008$ & $1.1078\pm0.0005$ \\
\bottomrule
\end{tabular}
\caption{Same as Table~\ref{tab:SI_semilep} for the dileptonic channel.}
\label{tab:SI}
\end{table}

\begin{table}[t]
\centering
\footnotesize
\setlength{\tabcolsep}{4pt}
\begin{tabular}{ll c cc cc cc}
\toprule
 & & $\Ssteer$ & \multicolumn{2}{c}{$139~\mathrm{fb}^{-1}$}
 & \multicolumn{2}{c}{$300~\mathrm{fb}^{-1}$}
 & \multicolumn{2}{c}{$3000~\mathrm{fb}^{-1}$} \\
\cmidrule(lr){4-5}\cmidrule(lr){6-7}\cmidrule(lr){8-9}
channel & method & (reconstructed) & precision & signif. & precision & signif.
 & precision & signif. \\
\midrule
\multicolumn{9}{l}{\emph{Boosted region}}\\
semileptonic & decay & $1.133\pm0.017$ & 1.5\% & $7.7\sigma$ & 1.0\% & $11\sigma$ & 0.33\% & $36\sigma$ \\
dileptonic & decay & $1.104\pm0.057$ & 5.2\% & $1.8\sigma$ & 3.5\% & $2.7\sigma$ & 1.1\% & $8.5\sigma$ \\
semileptonic & kinematic & $1.1334\pm0.0006$ & 0.052\% & $>100\sigma$ & 0.035\% & $>100\sigma$ & 0.011\% & $>100\sigma$ \\
dileptonic & kinematic & $1.1078\pm0.0005$ & 0.045\% & $>100\sigma$ & 0.031\% & $>100\sigma$ & 0.0097\% & $>100\sigma$ \\
\midrule
\multicolumn{9}{l}{\emph{Threshold region}}\\
semileptonic & decay & $1.18\pm0.12$ & 10\% & $1.5\sigma$ & 6.8\% & $2.2\sigma$ & 2.2\% & $7.1\sigma$ \\
dileptonic & decay & $1.18\pm0.24$ & 20.7\% & $0.7\sigma$ & 14.1\% & $1.1\sigma$ & 4.5\% & $3.4\sigma$ \\
semileptonic & kinematic & $1.2705\pm0.0002$ & 0.012\% & $>100\sigma$ & 0.0085\% & $>100\sigma$ & 0.0027\% & $>100\sigma$ \\
dileptonic & kinematic & $1.1647\pm0.0008$ & 0.066\% & $>100\sigma$ & 0.045\% & $>100\sigma$ & 0.014\% & $>100\sigma$ \\
\bottomrule
\end{tabular}
\caption{Precision $\delta\Ssteer/\Ssteer$ and significance
$(\Ssteer-1)/\delta\Ssteer$ for the measurement of the steerability integral
in the helicity basis after detector simulation and reconstruction, at
$\sqrt s=13$~TeV, for the current dataset and projected to Run~3 and the
HL-LHC assuming statistical scaling. The decay-method uncertainties are the
data statistics of the pseudo-experiments, as in Tables~\ref{tab:SI_semilep}
and \ref{tab:SI}.}
\label{tab:proj}
\end{table}

In the boosted region the fictitious state is
robustly steerable in both channels and with both methods, with
$\Ssteer\simeq1.11$--$1.14$ at parton level. The reconstruction does not shift
the central value appreciably, as expected for an unbiased unfolding. With the
decay method and the current dataset, steerability can be established at
$7.7\sigma$ in the semileptonic channel, whereas the dileptonic channel reaches
$1.8\sigma$ and requires about $1~\mathrm{ab}^{-1}$ for a $5\sigma$
observation. The factor of $3.3$ between the two uncertainties is larger than
the factor of $1.5$ expected from the branching fractions, spin analyzing
powers and reconstruction efficiencies alone~\cite{Han:2023fci}, reflecting the
stronger dilution of the measured asymmetries in the two-neutrino
reconstruction and the correspondingly larger unfolding uncertainty. By the
end of the HL-LHC the steerability integral will be measured to $0.3\%$
($1.1\%$) in the
semileptonic (dileptonic) channel with the decay method. As for quantum
discord~\cite{Han:2024ugl}, the kinematic method improves the precision by one
to two orders of magnitude, to about $0.05\%$ in both channels with current
data, because it does not require unfolding and in the boosted region the
kinematic variables are only mildly affected by detector effects. The price is that the result relies on the
leading-order templates for $C_{ij}(\theta,\Mtt)$ rather than on a direct
measurement of the spin correlations.

In the threshold region the optimized selection yields a fictitious state
that is steerable in both channels and with both methods at parton level,
$\Ssteer\simeq1.16$--$1.19$, with the decay-method values lying up to
$0.025$ below the kinematic-method ones, and the reconstruction preserves the
decay-method central values.    Establishing steerability at threshold with the decay method is, however,
   much harder than in the boosted region: near threshold the reconstructed mass
   is pushed upward, by about $20$~GeV ($35$~GeV) in the semileptonic
   (dileptonic) channel, so that a selection $15$--$40$~GeV wide is dominated by
   migration. Only about $40\%$ ($15\%$) of the selected events from the region
   are reconstructed inside it, while three quarters (two thirds) of the events
   reconstructed inside it originate from outside, so that the measurement relies
   on the simulated description of the migration and the uncertainty grows
   accordingly. With the current dataset the semileptonic
channel reaches $1.5\sigma$ (precision $10\%$) and the dileptonic channel
$0.7\sigma$; at the HL-LHC these become $7\sigma$ and $3.4\sigma$, and
$5\sigma$ is reached with about $1.5~\mathrm{ab}^{-1}$ in the semileptonic
channel. The threshold state is thus the second, independent region of phase
space in which EPR steering can be established, with a state of a different
nature, a singlet-dominated mixture close to the $\psi$ vertex of
Fig.~\ref{fig:face} rather than the triplet of the boosted region, but it is a
measurement for the HL-LHC. The kinematic method, which does not unfold, is
affected by the same migration: the reconstructed-level values in
Tables~\ref{tab:SI_semilep} and \ref{tab:SI} are shifted from the parton-level ones by far more than their statistical uncertainties, and that shift, rather
than the quoted uncertainty, sets the precision of the kinematic method in a
narrow region. 

Our analysis is performed in leading order in QCD and
neglects the background, which amounts to $\lesssim10\%$ of the signal after two
$b$-tags and are smaller in the boosted regime~\cite{Han:2023fci}.
Higher-order corrections are known to modify the spin correlations at the
$10$--$30\%$ level~\cite{Bernreuther:2004jv} and would shift the central values
of $\Ssteer$ accordingly; the impact of such corrections on quantum observables
has recently been studied for several other
processes~\cite{Chang:2026nzq,Goncalves:2025xer,Goncalves:2026njf,DelGratta:2025xjp,Pelliccioli:2026ltl}.
Near threshold, bound-state effects modify the spin correlations further and
have been proposed as a handle on toponium~\cite{Antozzi:2026vdi}; they are
particularly relevant for the threshold region. The convexity argument of Sec.~\ref{sec:fictitious}
does not rely on the leading-order form of the state, so the interpretation of
a measured $\Ssteer>1$ is unaffected as long as the top polarizations remain
negligible; the kinematic method, on the other hand, would need to be supplied
with higher-order templates. Finally, all projections assume statistically
dominated uncertainties; the systematic uncertainty from the unfolding is
included through the pseudo-experiments, but other experimental systematics
are not.

\section{Conclusions}
\label{sec:conclusions}

The top quark pair at the LHC has become a laboratory for quantum information
at the TeV scale. Following the observation of entanglement and the studies
of Bell nonlocality and quantum discord, we have presented a detailed
analysis of Einstein--Podolsky--Rosen steering in $\ttbar$ production, the
level of the hierarchy of quantum correlations that lies strictly between
entanglement and Bell nonlocality.

For the Bell-diagonal states produced at leading order, the steerability of
the $\ttbar$ state is decided by a single basis-independent number, the
steerability integral $\Ssteer$ of Eq.~\eqref{eq:SI}, which is twice the mean
length of the Bloch vector of one top quark conditioned on a spin measurement
of the other. We have derived elementary bounds on $\Ssteer$ in terms of the
singular values of the spin correlation matrix that provide a closed form sufficient condition $\sum_i\mu_i>3/2$, Eq.~(\ref{eq:suff}), and a closed form necessary 
condition $\mathrm{tr}(C^TC)>3/4$, Eq.~(\ref{eq:nec}), for steerability. We have obtained the steerability of the states $q\bar q$ in closed form as in Eq.~(\ref{eq:Sqq}) and $gg$ as in Eq.~(\ref{eq:ggwindow}) at $\theta=\pi/2$.
The $\ttbar$ state is steerable near the threshold $\Mtt\lesssim360$~GeV at central angles, and in
the boosted regime, $\Mtt\gtrsim800$~GeV, and these windows are nested between
the entangled and the nonlocal ones from Bell as the hierarchy requires. 

We develop for the first time the necessary formalism required to interpret steerability for fictitious states reconstructed at collider experiments.
Because
$\Ssteer$ is a convex function of the spin correlation matrix as discussed in Sec.~\ref{sec:fictitious}, a steerable
fictitious state reconstructed in the helicity basis guarantees the existence
of a steerable quantum substate, so that steering, unlike quantum discord, can
be measured at colliders without additional assumptions about the substates in a given region of phase space.
The same convexity implies that inclusive selections can hide steering, which
makes the choice of signal region an integral part of the measurement.

As presented in Sec.~\ref{sec:collider_analysis},
we have performed a full simulation of the semileptonic and dileptonic
channels, including parton shower, hadronization, detector response, kinematic
reconstruction and unfolding, and we have reconstructed the spin correlation
matrix with both the decay and the kinematic methods. In the boosted region,
steerability can be established with the decay method at $7.7\sigma$ in the
semileptonic channel with the current dataset, while the dileptonic channel
requires about $1~\mathrm{ab}^{-1}$ to reach $5\sigma$. The steerability
integral will be measured to $0.3\%$ ($1.1\%$) at the HL-LHC in the
semileptonic (dileptonic) channel, and the kinematic method improves the
precision by one to two orders of magnitude. This region is steerable but not
Bell nonlocal, so that steering is the strongest form of quantum correlation
accessible there with present data. The inclusive threshold region used in
entanglement and discord studies, on the other hand, is firmly entangled but
lies at the steerability boundary; an optimized selection,
$\Mtt\le360$~GeV supplemented by two forward wedges, isolates a steerable
singlet-dominated state. Its observation with the decay method is limited by
the migration across the narrow selection and requires the HL-LHC, where
steering at threshold can be established at $7\sigma$ ($3.4\sigma$) in the
semileptonic (dileptonic) channel; with current data the semileptonic channel
reaches $1.5\sigma$.

Several directions remain open. At leading order the steering of $\ttbar$ is
symmetric; one-way steering requires top polarization and is therefore a probe
of electroweak corrections and of new
physics~\cite{Aoude:2025jzc,Belmonte:2026yno}, in close analogy with the
asymmetry of quantum discord as a signal of CP violation, while at lepton
colliders beam polarization offers a direct handle on the polarizations and
hence on steering~\cite{Zhang:2026wvn,Altakach:2026fpl,Guo:2026yhz}. The impact of
higher-order QCD corrections and of backgrounds on $\Ssteer$, the extension to
the fully hadronic channel, and the use of steering inequalities with a finite
number of settings as experimentally simpler, if weaker, witnesses are natural
next steps. More broadly, the methods developed here apply to any bipartite
qubit system produced at colliders, and the closed-form criteria of
Sec.~\ref{sec:steer-bounds} should prove useful for
$\tau^+\tau^-$~\cite{Zhang:2025mmm,Afik:2026api,Zhang:2026wvn} and other final states, and
complement the geometric characterization of steering at colliders of
Ref.~\cite{Alvarez:2026uve}.

\acknowledgments

We acknowledge the use of Claude (Anthropic) for assistance with text editing, figure generation, and coding, and ChatGPT (OpenAI) for coding and cross checks. The authors take full responsibility for the accuracy of the content. TH is supported in part by the US Department of Energy under grant No.~DE-SC0007914, and in part by Pitt PACC. The work of NM and SS has been supported by the U.S.~Department of Energy under grant No.~DEFG02-13ER41976/DE-SC0009913.  NM is also supported by  the DOE QuantISED program through the theory consortium ``Intersections of QIS and Theoretical Particle Physics'' at Fermilab (FNAL 20-17) under contract No.~89243024CSC000002.

\bibliographystyle{JHEP}
\bibliography{steering_refs}

\end{document}